\documentclass[a4paper,11pt]{article}
\usepackage{a4wide}
\usepackage{epsfig}

\newcommand{\mc}[1]{$\mathcal{#1}$}

\title{Required experimental accuracy to select between 
supersymmetrical models}

\author{David Grellscheid\\
{\small DAMTP, CMS,  University of Cambridge,}\\
{\small Cambridge CB3 0WA, UK}}

\begin{document}
\maketitle
\begin{abstract}{%
We will present a 
method to decide a priori
whether various supersymmetrical scenarios can be distinguished based on
sparticle mass data alone. For each model, a scan over all free SUSY
breaking parameters reveals the extent of that model's physically allowed
region of sparticle-mass-space. Based on the geometrical configuration of
these regions in mass-space, it is possible to obtain an estimate of the
required accuracy of future sparticle mass measurements to distinguish
between the models. We will illustrate this algorithm with an example.
This talk is based on work done in collaboration with B.C.~Allanach
(LAPTH, Annecy) and F.~Quevedo (DAMTP, Cambridge).}
\end{abstract}

\section*{Background} 
Rather than analyzing single mSUGRA points in detail
\cite{atlastdr} or reconstructing
SUSY breaking parameters from low scale observables \cite{zerwas}, 
we propose a procedure aiming to 
look at various models of SUSY breaking simultaneously, scanning over wide
ranges of their input parameters to try to find the minimal set of
measurements that is required to separate the models.
We wish to investigate whether it is possible to decide a priori that
two high scale models can be distinguished on experimental grounds,
and want to determine the necessary measurement accuracy to do so.

At the root of our procedure stands the observation that
any sparticle spectrum study basically deals
with two kinds of quantities. On the one hand
we have a set of input parameters at the high scale, determined by some
fundamental theory; on the other we have a set of
observables at the electroweak scale. Both sets of quantities can be
looked at as vectors in their respective parameter spaces.
The first space, \mc{I}, is one of free model
parameters at the high scale. Each model $m$ under
consideration will have its own input space \mc{I}${}_m$. The number of its
dimensions 
is determined by the number of free parameters in the
model.
To make our analysis technically 
feasible, this should be a
small number, typically smaller than 6--8. Each point
in \mc{I}${}_m$ then corresponds to one fixed choice of high-scale input
parameter values for model $m$.

The second space, \mc{M}, is the space of physical
measurements at the electroweak scale. There is only one
unique \mc{M}, since all models describe the same
electroweak scale physics. 
 Its dimensionality
equals the number of
low-scale observables under consideration. Typical values are as
large as 20--30 (taking in all sparticle masses). 
Each point in \mc{M} denotes one set of fixed values for the observables. 

\begin{figure}[htbp]
\centerline{\epsfig{file=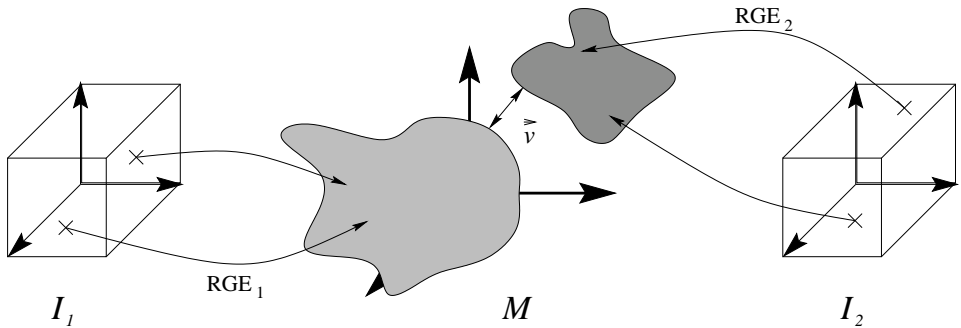, width=5in}}
\caption{\mc{I}${}_1$ and \mc{I}${}_2$ are input parameter spaces for
two different SUSY breaking models.
Each point within \mc{I}${}_m$ corresponds to one set of high-scale
parameters for model $m$, serving as input to this model's RGEs. 
They uniquely map each input
point onto a point in \mc{M}, the space of measurements. Scanning over
\mc{I}${}_m$ point by point builds up the footprint of model $m$ in
\mc{M}.
The closest approach of the two footprints is indicated by $\vec{v}$.}
\label{map}
\end{figure}

Each model also specifies a set of renormalization group equations
(often this may be the standard MSSM RGEs), through which
each point in \mc{I}${}_m$ 
can potentially be mapped onto a point in \mc{M} (see figure \ref{map}).
Consequently, a scan over all $N$ parameters in
\mc{I}${}_m$ will build up an  $N$-dimensional hypersurface
in \mc{M}, which we will call the \emph{footprint} of the high-scale
model under consideration. One restriction is imposed at this step:
only physical points which do not violate experimental 
bounds are considered part of the footprint. 

Different models will have different
footprints, some of which may be disjoint, while 
others may overlap. 
However, as long as the footprints' hypersurfaces 
are of much lower dimensionality
than \mc{M}, as is generally the case, it will be quite unlikely
that there will be any overlap between the prints. 
As soon as it is established that the two prints are
disjoint, it is  possible to conclude that the two models can in
principle be distinguished experimentally, as long as a certain
 measurement accuracy can be achieved.
To determine the required level of accuracy,
a minimization algorithm can
be used to find $\vec{v}$, the vector 
spanning the closest approach of the two footprints.
The required relative accuracy for the various measurements (which lie
along the axes of 
\mc{M}) can now be determined by examining
the respective components of $\vec{v}$.

\section*{Example study}
As a test case for our procedure we looked at the three Type I String
motivated models we have used previously in \cite{agq}. The input spaces
here are four-dimensional subsets of the standard mSUGRA parameter
space, and they all
use the same set of parameters (namely two goldstino angles
$\theta$ and $\phi$, $\tan\beta$ and the gravitino mass $m_{3/2}$). 
The models differ in the running of the
RGEs that were used to obtain the footprints. 
One model assumes the
standard gauge coupling unification at
$m_{GUT}=10^{16}\,\mathrm{GeV}$, 
the second uses a fundamental scale of $10^{11}\,\mathrm{GeV}$ with mirage
unification, 
the third achieves early unification at
$10^{11}\,\mathrm{GeV}$ through the addition of extra slepton
multiplets (see \cite{agq} for details).
To obtain the sparticle spectra, we used a slightly modified version of 
\textsc{Softsusy 1.7.1} \cite{softsusy}.

The determination of $\vec{v}$
turned out to be problematic for standard minimization algorithms
such as \textsc{Minuit} \cite{minuit} because of the very
irregular nature of the footprints' boundaries which originates in the
exclusion of unphysical points. A more promising
approach which we are currently working on is
 the use of Genetic Algorithms \cite{GA}, which are more
robust against the occurrence of excluded points. 
The choice of observables to be plotted against the axes in \mc{M}
turned out to be
somewhat  critical. We found it to be useful to plot mass ratios
rather  than the masses directly, to eliminate the dependence of the
plots on the overall size of the SUSY mass splittings.
\begin{figure}[htbp]
\centerline{\epsfig{file=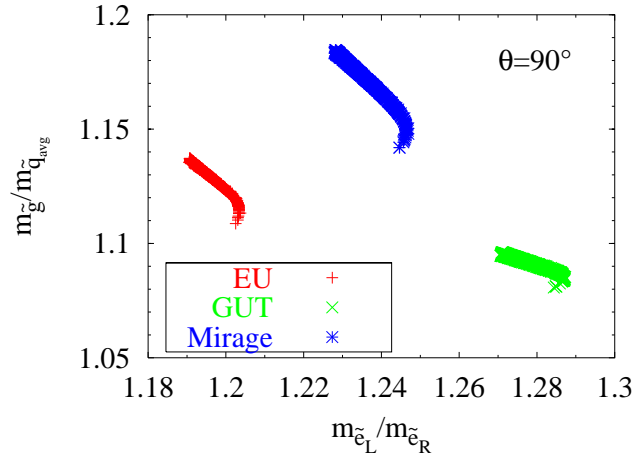, width=3.5in}}
\caption{%
Footprints generated by scanning 1000 random points in the
dilaton domination limit
for each of the three considered models. (Updated figure from [3])
}
\label{agqplot}
\end{figure}

In \cite{agq}, we have shown that the separation of the three models
is possible in the dilaton dominated case (see figure \ref{agqplot}),
where one of the input parameters is held at a fixed value: 
$\theta = 90^\circ$. To distinguish the models, a combined theoretical and
experimental accuracy \cite{newben} of about 2--3\% is
required for the determination of both
$m_{\tilde{g}}/m_{\tilde{q}_{avg}}$ and
$m_{\tilde{e}_L}/m_{\tilde{e}_R}$, which could be achievable in a
combined linear collider / LHC analysis. 

With our new automatized approach, we were able to extend this
result to show separation for $60^\circ \leq \theta\leq90^\circ$, 
and are currently working on extending it towards the full
dimensionality scan over all four free input parameters.

Possible future directions for this work 
could include a study of other high scale
models, which do not necessarily have to be string models. Large
improvements can  probably be made in the design of the
minimization algorithm and the choice of investigated observables.

\end{document}